\documentclass[aps,prb,twocolumn]{revtex4}
\usepackage{psfig}
\begin{document}
\draft

\title{The Thermal Conductivity Reduction in HgTe/CdTe Superlattices}
\author{W. E. Bies and H. Ehrenreich}
\address{Physics Department and
Division of Engineering and Applied Sciences,\\
Harvard University, Cambridge, Massachusetts~~02138}
\author{E. Runge}
\address{
AG Halbleitertheorie, Inst. Physik, Humboldt-Universitaet zu Berlin\\
Hausvogteiplatz 5-7, 10117 Berlin, Germany}
\date{December 2, 2001}

\begin{abstract}
The techniques used previously to calculate the three-fold thermal
conductivity reduction due to phonon dispersion in GaAs/AlAs
superlattices (SLs) are applied to HgTe/CdTe SLs. The reduction
factor is approximately the same, indicating that this SL may be
applicable both as a photodetector and a thermoelectric cooler.
\end{abstract}
\maketitle

\newpage

\section{Introduction}

Hg$_x$Cd$_{1-x}$Te alloys are among the key materials used for cooled infra-red detectors.  Materials structures involving these constituents in both alloy and superlattice (SL) forms that would be useful for both optical detection and thermoelectric cooling to a reasonably low temperature are an attractive possibility.  Some semiconducting superlattices (SLs) are known to exhibit an order-of-magnitude reduction in the lattice thermal  conductivities\cite{HD,R1,R2,R3,HM,TTM,CN,MW,MSB,RHG}. For example, Capinksi {\it et al} have experimentally shown GaAs/AlAs SLs to have a ten-fold smaller lattice thermal conductivity than the bulk alloys having the same composition\cite{capinski1,capinski2}. This result implies that the thermoelectric figure of merit ZT would be increased by an appreciable factor.  We show theoretically that the same appears to be true for HgTe/CdTe SLs.  Cooling of integrated room temperature devices to 200-250 K thus becomes a possibility for applications requiring no external coolant provided these temperatures are adequate.  Since the devices would be fabricated of the same material constituents, manufacturing costs might be substantially reduced.

We have previously investigated GaAs/AlAs theoretically \cite{BRE} using Kunc's model for the lattice spectrum \cite{kunc1,kunc2} generalized to SLs\cite{RCC}. Details are  given in Ref. 13.  This rigid-ion model considers next-nearest neighbor short-range forces together with the long-range Coulomb force.  We found a factor-of-three reduction in the lattice thermal conductivity along the growth direction due to SL induced modifications of the phonon spectrum.  The remainder of the overall ten-fold reduction is presumably associated with phonon scattering at the interfaces \cite{BRE,chen1,chen2}  which is not considered in the present work.  The thermal conductivity reduction in question here can be interpreted as resulting from zone folding in the SL relative to bulk. The new gaps introduced at the zone face result in band flattening and thus lower phonon velocities, which in turn reduce the thermal conduction.  These effects are illustrated in Fig. 1 and Fig. 4 of Ref. 13.  The corresponding figures in this paper will be seen to exhibit very similar effects in HgTe/CdTe SLs.  For reasons to be discussed, these results  lead us to expect similar behavior in any SL having the same structure and similar mass and force constant differences between layers.  Due to residual ``impedance mismatches" at the interfaces, the Kunc model leads to decaying TA and LA modes along some intervals and directions in q-space.  (This is not the case for GaAs/AlAs SLs.)  We resolve this problem by use of a variety of physically motivated modifications, which are critically examined in the next section. The same three-fold reduction previously calculated for GaAs/AlAs is obtained.  Moreover the result is shown to be robust with respect to the types of physically reasonable approximations made.  Our discussion emphasizes that a reasonable phonon spectrum is obtained if one of the force constants, obtained from neutron diffraction experiments for HgTe and CdTe respectively, is modified by a small amount. Unfortunately there are no corresponding neutron data for the SL that might lead to better experimental input for determining the Kunc force constants.

\section{The Kunc model for H\lowercase{g}T\lowercase{e}/C\lowercase{d}T\lowercase{e}
SL\lowercase{s}}

The Kunc rigid-ion model \cite{kunc1,kunc2} has been successfully
applied to compounds with zinc-blende structure. In each bulk compound,
ten parameters describe the force constants between nearest and next-nearest
neighbors and one parameter the effective charge. In the absence of
experimental data for HgTe/CdTe SLs there are no directly
determinable parameters with which to fit the phonon spectrum.
Note that even for bulk HgTe the fit to the neutron-diffraction data
along the [111] direction does not reproduce the optical branches 
very well \cite{TV}. 
Thus, we use the parameter fits to bulk HgTe and
CdTe determined by Talwar and Vandevyver \cite{TV} 
and construct the SL dynamical matrix from these. There are thus 22
parameters needed for the SL, in addition to the atomic masses and
lattice spacing. The bulk parameters are used within each layer. At the
interface we average the force constants from the two sides. The Madelung sum is calculated using the Ewald transformation adapted
to the SL. We refer to Ref.~\onlinecite{BRE} for the details. The output of the
Kunc model is the phonon dispersion $\omega_{\bf q}^{(\alpha)}$ for
the $\alpha$-th branch as a function of wavevector ${\bf q}$ in the
SL Brillouin zone. The lattice thermal conductivity may then be computed
from the phonon Boltzmann equation in the relaxation-time approximation:
\begin{eqnarray}
(\kappa_{\ell})_{ij} &=&
\int {{d^3q} \over {(2\pi)^3}} \left[ \sum_\alpha
\hbar \omega^{(\alpha)}_{\bf q} {{\partial \omega^{(\alpha)}_{\bf q}} \over {\partial q_i}}
{{\partial \omega^{(\alpha)}_{\bf q}} \over {\partial q_j}} {{dn(\omega^{(\alpha)}_{\bf q})} \over dT} \right] \times \nonumber \\
&& \tau_{\rm ph}(\omega^{(\alpha)}_{\bf q},T),
\label{def_kappa}
\end{eqnarray}
where $n(\omega^{(\alpha)}_{\bf q})$ is the distribution function
of phonons and {$\tau_{\rm ph}(\omega_{\bf q}^{(\alpha)},T)$} is the lifetime,
taken to be a constant $\tau$.

The dynamical matrix for the HgTe/CdTe SL so constructed, though Hermitian,
has negative eigenvalues, i.e., yields imaginary phonon frequencies, for the three acoustic branches in a cylindrical region of the Brillouin zone
around the [001] axis that occupies about 6\% of the zone by volume.
The problem of imaginary frequencies in models of HgTe/CdTe SLs has been
investigated by Rajput and Browne.\cite{RB} They attribute the problem
to an inadequate modelling of the interface, in their simple force model based
on the Keating potential. They find that in their model the imaginary frequencies can be removed by treating the two materials as continuum media with differing dielectric constants. However, this prescription fails to remove the imaginary frequencies using the Kunc model. As we shall see below, the imaginary frequencies
in the Kunc model are more likely to arise from the force constants than from the Coulomb interaction.

In order to gain an idea of what is causing the imaginary frequencies,
we looked at the dependence of the gap at the zone edge, $\Delta \omega$,
on the Kunc parameters. We found the strongest dependence for the
diagonal terms in the force constant matrices; the effect of the
off-diagonal terms was weaker by a factor of ten or more.
This finding motivated us to consider the variation of the parameters necessary to remove imaginary frequencies.
For example, an interpolation of all parameters (20 force constants,
2 effective charges, 3 atomic masses and the lattice spacing)
linearly between between their values for a HgTe/CdTe SL ($x=0$) and
those for a GaAs/AlAs ($x=1$), which has real frequencies
only, shows that all of the imaginary frequencies disappear when $x$ approaches 0.035.
This suggests that we associate the imaginary frequencies with an impedance mismatch between the HgTe and CdTe layers which can be
removed by variation of appropriate force constants or the introduction
of additional ones provided by a more sophisticated model.
Thus, if we hold the parameters fixed at their HgTe/CdTe values and
vary the parameters individually, we find that small ($<10$\%) changes in $A_{\rm HgTe}$ and $A_{\rm CdTe}$ alone are sufficient to yield real
frequencies throughout the SL Brillouin zone. Here, $A$ is the nearest-neighbor coupling between Hg and Te respectively Cd and Te. Apparently the problem does not exist in
GaAs/AlAs SLs because the $A$ parameters are so closely matched:
$A_{\rm GaAs}=-40.77$ N/m and $A_{\rm AlAs}=-40.83$ N/m.

We now describe our approaches for removing the imaginary
frequencies. The only correct way to solve the problem is to
generalize the already complicated Kunc model. However, the Kunc-model
description of the bulk phonon dispersion is satisfactory. Changing the bulk model does not necessarily imply that the SL results will be
improved because the interfaces are neglected. Instead, we construct physical models, for example, determining values of  $A$  that eliminate the imaginary frequencies. The reduction factor is always about 3 for the more realistic
models. The robustness of the result gives us more confidence in its
physical relevance than would deriving it from a generalized Kunc model
with many new parameters.

The best model utilizes scaling, which maintains the acoustical
and optical mode frequencies observed in bulk.
The effect of changing $A$, aside from removing the imaginary frequencies,
is to rescale all the phonon frequencies across the Brillouin zone.
The scaling model is therefore constructed as follows: (1) $A$ is
replaced by $A'$; in practice $A'$ is selected so that the ratio of
LA to TA sound velocities in the [001] direction will be that of bulk;
this means $x=0.2$ or $A_{\rm HgTe}$ is changed from $-18.127$ N/m to
$-22.7$ N/m and $A_{\rm CdTe}$ from $-19.97$ N/m to $-24.1$ N/m. This gives
sound velocities about 10\% too high and increases the frequencies of the
optical branches by 10\%. (2) In order to compensate for this increase,
all phonon frequencies are rescaled by a common factor,
$\omega \rightarrow \lambda \omega$, with $\lambda=0.927$, assuming
$\lambda$ to be constant. The scaling model has the advantage of
staying within the Kunc-model framework, while removing the imaginary frequencies in a consistent way.

Another model is based on averaging the force constants (both nearest neighbor
and next-nearest neighbor) at the interface. This approach serves to equalize the force experienced in each direction by the
interfacial Te ion. This approach fails, however, to remove the
imaginary frequencies. The fraction of the Brillouin zone covered
by imaginary frequencies changes by less than 1\%.

The problem can be addressed most simply by setting the frequencies of the imaginary branches equal to zero. We find a thermal conductivity
reduction by a factor of 3.0, very close to that found for GaAs/AlAs.
Since only three of the eighteen folded-back acoustic branches are
set to zero over a small part of the zone, the end result should be
close to the true value, which represents a lower bound to the reduction factor.
For comparison, we find that setting the lowest three branches to zero
in GaAs/AlAs gives a reduction by a factor of 2.9, versus 2.8 in the
correct calculation. Thus we can expect that the actual value in HgTe/CdTe
will be close to 3.

Another approach involves a linear interpolation from the origin to the point of tangency with the acoustic
branches where they are real. This gives a reduction factor of 2.9.
The treatment of the acoustic branches however is unrealistic because
the interpolated sound velocities are too low, only about 50\% of the
bulk value (average of HgTe and CdTe) in the [001] direction.

The simplest model is based on the virtual crystal approximation, in which
the force constants are replaced with their averages between the two
layers, while the masses and effective charges are kept at their
respective values for the two layers. This ensures that
the frequencies are real, because of the absence of interfaces.
However, it does not properly account for the thermal conductivity reduction (giving only a factor of 1.3), because the band gaps at the zone edge, which reduce the phonon velocity, are
considerably smaller than they should be. 
These results indicate that it is the force constants
and not the Coulomb interaction that are responsible for the imaginary frequencies
in the SL Kunc model. 
\section{Results}

We consider a (HgTe)$_3$/(CdTe)$_3$ SL. The dispersion relation along the $\Gamma X$ and $\Gamma Z$ directions was generated numerically
using the scaling model as described in Sec.~II and is shown in
Fig.~1. The SL unit cell contains three unit cells each of
HgTe and CdTe, arranged along the growth axis. Thus, the edge of the
SL Brillouin zone is one-sixth as far from the center along the $\Gamma Z$
axis as it is in the in-plane $\Gamma X$ and $\Gamma Y$ directions.
Each of the six branches in bulk is folded back six times in the
$\Gamma Z$ direction, yielding 36 branches in the SL. This effect
is most easily seen for the acoustic modes along $\Gamma Z$.
The results are qualitatively similar to those obtained for
(GaAs)$_3$/(AlAs)$_3$ in Ref.~\onlinecite{BRE}. The HgTe and CdTe optical
modes are not as well separated as the GaAs and AlAs optical modes are in
GaAs/AlAs, and we see that the lowest HgTe optical modes overlap with the highest acoustic modes along $\Gamma X$. Nevertheless, the optical
modes are nearly flat in the $\Gamma Z$ direction.

As can be seen in Fig.~1, the physical effects of band flattening,
relative to bulk, are present. We therefore expect, by Eq.~(\ref{def_kappa}),
a reduction in the thermal conductivity along the growth axis.
In fact, with our data, Eq.~(\ref{def_kappa}) yields a reduction factor
in $\kappa/\tau$ of 3.0 (in the scaling model). This factor reflects
phonon-dispersion effects only. As discussed in Ref.~\onlinecite{BRE},
one may expect a further reduction in thermal conductivity if the lifetime $\tau$ itself is lowered in the SL compared to bulk due primarily to interface scattering. But there is not sufficient data
on HgTe/CdTe SLs for us to discuss the lifetime quantitatively.

A simple physical picture of the thermal conductivity reduction due to phonon dispersion effects is obtained by looking at the
transport quantity {$q_x \sum_\alpha (dn(\omega^{(\alpha)}_{\bf q})/dT) \omega^{(\alpha)}_{\bf q} (v^{(\alpha)}_{{\bf q},z})^2$} in {\bf q} space.
This is the quantity in square brackets in
Eq.~(\ref{def_kappa}). Since the dispersion
relation is rotationally symmetric in the $(q_x,q_y)$ plane 
to good approximation we perform an
annular integration, yielding the factor $q_x$ equal to the annular
radius and leaving an integral in Eq.~(\ref{def_kappa}) depending
only on $q_x$ and $q_z$. The resulting quantity is
plotted in Figs.~2(a), 2(b) and 2(c) for the SL and for
bulk HgTe as a function of $q_z$ for three values
of $q_x$. The SL transport quantity is seen to be localized, i.e., it
tends to zero at the band edges $q_z=0$ and $q_z=\pi/d$. The reduction
in the SL transport quantity near the zone edges is related to
miniband formation (the flattening of $\omega$ versus $q_z$ seen in
Fig.~1). To gain physical insight, we replace the SL transport 
quantity in Figs.~2(a), (b) and (c)
by rectangles of equal area and plot these in Fig.2(d), which summarizes
the dependence of the transport quantity on $q_x$. The shading indicates
the weight of each equivalent rectangle within the range $(q_x,q_x+\Delta q_x)$. In bulk the equivalent rectangles extend all the
way to the zone edges because there is no localization due to band
flattening. In Fig.~2(d) we see that the densities tend to zero as
$q_x$ goes to zero and that the the density in bulk increases for
large $q_x$, again due to the $q_x$ prefactor. Surprisingly, the
density in the SL falls off fast enough as $q_x$ increases to overcome
the $q_x$ factor. This should be compared with the results for GaAs/AlAs
SLs of Ref.~\onlinecite{BRE}.
The qualitative difference between HgTe/CdTe
and GaAs/AlAs SLs is not apparent from their dispersion relations.
Integrating the density in Fig.~2 gives an estimated
reduction factor of 2.4, which is to be compared with 3.0 in the
exact calculation. The difference is due to the fact that the estimate
does not properly treat the region of the Brillioun zone with radius
greater than $\sqrt{2}\pi/a_0$ and therefore underestimates the reduction
factor.

Our results predict a reduction factor in thermal conductivity for HgTe/CdTe
SLs comparable in size to those already observed for GaAs/AlAs SLs. The minimum cold temperature achievable for a 30\AA ~Hg$_{0.75}$Cd$_{0.25}$Te/30\AA ~ Hg$_{0.7}$Cd$_{0.3}$Te SL is about 180 K, compared to 240 K for conventional Bi$_2$Te$_3$.  This result suggests that it is reasonable to
use HgCdTe for thermoelectric
cooling applications in infra-red detectors.

\section{Acknowledgments}

The authors wish to thank C.H. Grein for helpful discussions.  This work was supported by DARPA through ARL Contract No. DAAD17-00-C-0134.

\begin{figure}

\centerline{
\psfig{file=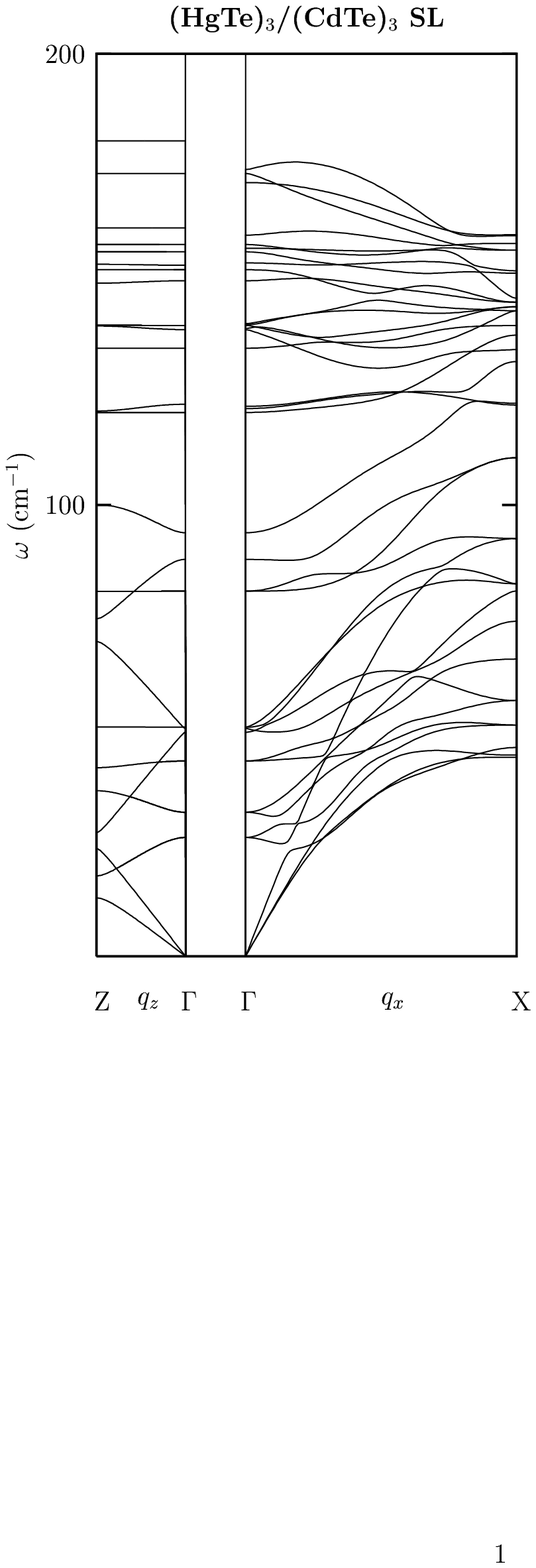,width=3.5in,bbllx=85pt,bblly=270pt,bburx=325pt,bbury=720pt,clip=}}

\caption{
(HgTe)$_3$/(CdTe){$_3$} superlattice dispersion relation along the
{$\Gamma$}X={$(2\pi/a_0,0,0)$} and {$\Gamma$}Z={$(0,0,\pi/3a_0)$} directions,
where {$a_0=6.48$} {\AA} is the conventional unit cell size of bulk HgTe.}

\end{figure}

\newpage

\begin{figure}

\centerline{
\psfig{file=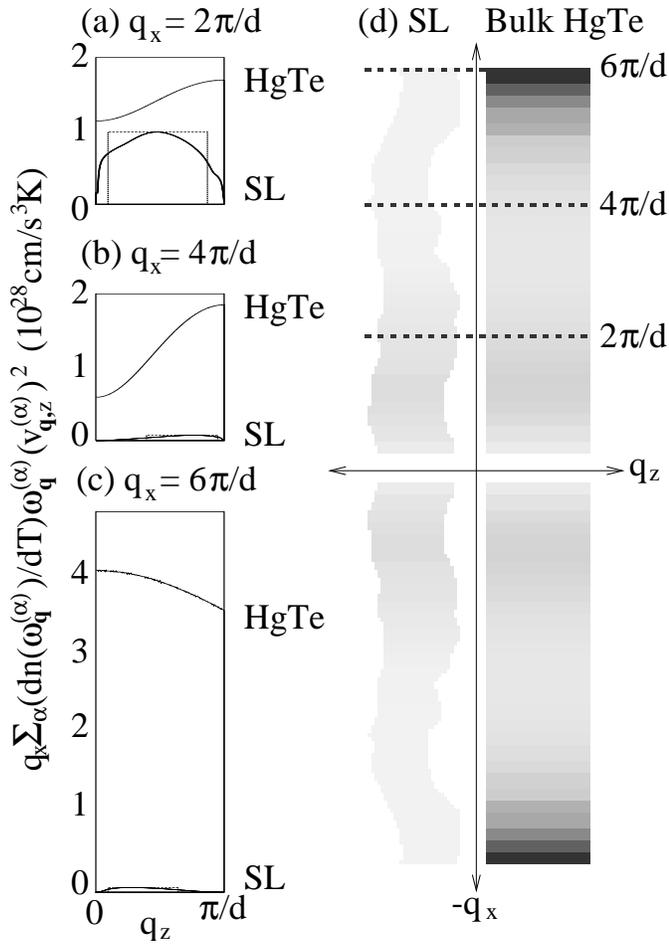,width=3.5in,bbllx=60pt,bblly=50pt,bburx=555pt,bbury=750pt,clip=}}

\caption{(a),(b),(c) The transport quantity {$q_x \sum_\alpha (dn/dT) \omega_{\bf q}^{(\alpha)} (v_{{\bf q},z}^{(\alpha)})^2$}
for {$0 \le q_z \le \pi/d$} at fixed {$q_x$}
for (a) {$q_x=2\pi/d$}, (b) {$q_x=4\pi/d$} and (c) {$q_x=6\pi/d$}.
The bulk HgTe and the (HgTe)$_3$/(CdTe){$_3$} SL curves are
labelled on the right of figures (a), (b), and (c). Note the smallness
of the SL curves in (b) and (c). (d) Density plot in the {$(q_x,q_z)$}
plane whose shading indicates the weight of each increment in
$\Delta q_x$ along $q_x$ to the value of the transport quantity
for the (HgTe)$_3$/(CdTe){$_3$} SL and bulk HgTe respectively.}

\end{figure}

\end{document}